\documentclass[sigconf]{acmart}
\renewcommand\footnotetextcopyrightpermission[1]{} % removes footnote with conference information in first column
\def\BibTeX{{\rm B\kern-.05em{\sc i\kern-.025em b}\kern-.08emT\kern-.1667em\lower.7ex\hbox{E}\kern-.125emX}}

\copyrightyear{2019} 
\acmYear{2019} 
\acmConference[NSPW '19]{New Security Paradigms Workshop}{September 23--26, 2019}{San Carlos, Costa Rica}
\acmBooktitle{New Security Paradigms Workshop (NSPW '19), September 23--26, 2019, San Carlos, Costa Rica}
\acmPrice{15.00}
\acmDOI{10.1145/3368860.3368866}
\acmISBN{978-1-4503-7647-1/19/09}

\usepackage{graphicx}% http://ctan.org/pkg/graphicx
\usepackage{listings}
\usepackage{xcolor}
\usepackage{balance}

\definecolor{codegreen}{rgb}{0,0.6,0}
\definecolor{codegray}{rgb}{0.5,0.5,0.5}
\definecolor{codepurple}{rgb}{0.58,0,0.82}
\definecolor{backcolour}{rgb}{0.95,0.95,0.92}
\lstdefinestyle{mystyle}{
  backgroundcolor=\color{backcolour},   commentstyle=\color{codegreen},
  keywordstyle=\color{magenta},
  numberstyle=\tiny\color{codegray},
  stringstyle=\color{codepurple},
  basicstyle=\ttfamily\footnotesize,
  breakatwhitespace=false,         
  breaklines=true,                 
  captionpos=b,                    
  keepspaces=true,                 
  numbers=left,                    
  numbersep=5pt,                  
  showspaces=false,                
  showstringspaces=false,
  showtabs=false,                  
  tabsize=2
}
\begin{document}

\title{FrameProv: Towards End-To-End Video Provenance}

\author{Mansoor Ahmed-Rengers}
\email{mansoor.ahmed AT cl.cam.ac.uk}
\affiliation{%
  \institution{Department of Computer Science and Technology}
  \city{University of Cambridge}
}

\begin{abstract}
Video feeds are often deliberately used as evidence, as in the case of CCTV footage; but more often than not, the existence of footage of a supposed event is \emph{perceived} as proof of fact in the eyes of the public at large. This reliance represents a societal vulnerability given the existence of easy-to-use editing tools and means to fabricate entire video feeds using machine learning. And, as the recent barrage of fake news and fake porn videos have shown, this isn't merely an academic concern, it is actively been exploited. I posit that this exploitation is only going to get more insidious. In this position paper, I introduce a long term project that aims to mitigate some of the most egregious forms of manipulation by embedding trustworthy components in the video transmission chain. Unlike earlier works, I am not aiming to do tamper detection or other forms of forensics -- approaches I think are bound to fail in the face of the reality of necessary editing and compression -- instead, the aim here is to provide a way for the video publisher to prove the integrity of the video feed as well as make explicit any edits they may have performed. To do this, I present a novel data structure, a video-edit specification language and supporting infrastructure that provides end-to-end video provenance, from the camera sensor to the viewer. I have implemented a prototype of this system and am in talks with journalists and video editors to discuss the best ways forward with introducing this idea to the mainstream.
\end{abstract}


%
% Keywords. The author(s) should pick words that accurately describe the work being
% presented. Separate the keywords with commas.
\keywords{video provenance, video editing, tamper evidence}

%
% This command processes the author and affiliation and title information and builds
% the first part of the formatted document.

\begin{CCSXML}
<ccs2012>
<concept>
<concept_id>10002978.10003006.10003007.10003009</concept_id>
<concept_desc>Security and privacy~Trusted computing</concept_desc>
<concept_significance>500</concept_significance>
</concept>
<concept>
<concept_id>10002978.10003029.10011703</concept_id>
<concept_desc>Security and privacy~Usability in security and privacy</concept_desc>
<concept_significance>300</concept_significance>
</concept>
<concept>
<concept_id>10010405.10010462</concept_id>
<concept_desc>Applied computing~Computer forensics</concept_desc>
<concept_significance>300</concept_significance>
</concept>
<concept>
<concept_id>10003120.10003130.10003131.10003234</concept_id>
<concept_desc>Human-centered computing~Social content sharing</concept_desc>
<concept_significance>100</concept_significance>
</concept>
</ccs2012>
\end{CCSXML}

\ccsdesc[500]{Security and privacy~Trusted computing}
\ccsdesc[300]{Security and privacy~Usability in security and privacy}
\ccsdesc[300]{Applied computing~Computer forensics}
\ccsdesc[100]{Human-centered computing~Social content sharing}

\maketitle

\section{Introduction}

That fake news is a tricky problem to solve, is probably not news to anyone at this point. However, it is my belief that this problem stands to get a lot trickier once the fakesters open their eyes to the potential of a mostly untapped weapon: trust in videos. Fake news so far has relied on social media segregation and textual misinformation with the odd photoshopped picture thrown here and there. This has meant that, by and large, an intellectually honest curious individual has been able to discover the truth behind the fakeness. This could soon change. You see, ``pics or it didn't happen'' isn't just a meme, it is the mental model by which people judge the veracity of a piece of information on the Internet. What happens when the fakesters are able to create forgeries that even a keen eye cannot distinguish? How does a curious person distinguish truth from fact?

We are far closer to this future than many realize: in 2017, researchers created a tool that produced realistic looking video clips of Barack Obama saying things he has never been recorded saying~\cite{obama}. Since then, a barrage of similar tools have become available; an equally worrying, if slightly tangential, trend is the rise of fake pornographic video that superimpose images of celebrities on to adult videos. These tools represent the latest weapons in the arsenal of fake news creators -- ones far easier to use for the layman than those before. While the videos produced by these tools may not presently stand up to scrutiny by forensics experts, they are already good enough to fool a casual viewer (and getting better). The end result is that creating a good-enough fake video is now a trivial matter.

There are, of course, more traditional ways of creating fake videos as well. The white house itself was caught using the oldest trick in the book while trying to justify the barring of a reporter from the briefing room: they sped up the video to make it look like the reporter was physically rough with a staff member~\cite{whitehouse}. Other traditional ways are misleadingly editing videos to leave out critical context~\cite{plannedparenthood}, or splicing video clips to map wrong answers to questions, etc. I expect that we will see a strong increase in these traditional fake videos before we see a further transition to the complete fabrications discussed above. Both represent a grave danger to the spread of truth.

\begin{figure*}
  \includegraphics[width=\textwidth]{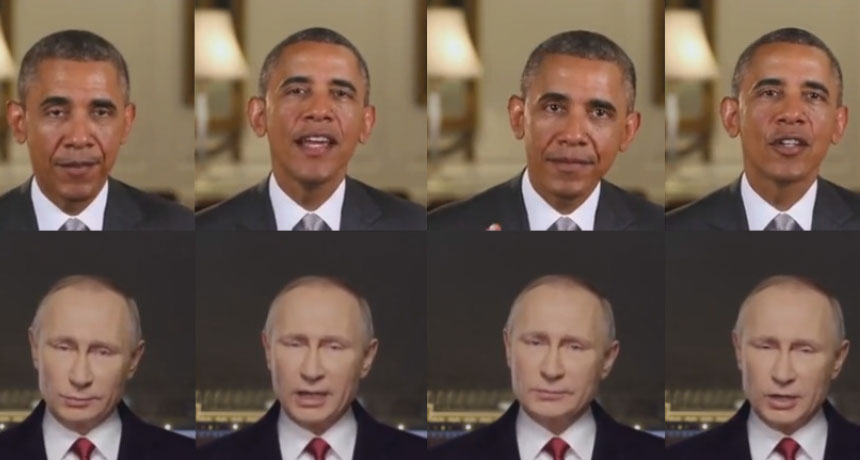}
  \caption{Did Obama say that or Putin? Deep fake videos such as this (where researchers produced an Obama speech he never said) are right around the corner. We are ill-prepared to deal with this form of supercharged fake news; FrameProv offers a way to mitigate these, and other, kinds of forgeries.}
  \label{fig:teaser}
\end{figure*}

The consequences of fake news have been grave in western countries, but their effects have undoubtedly been far  deadlier in less technologically literate areas of the world. Fake news spread via whatsapp has led to countless murders and riots in the last couple of years in India~\cite{whatsappindia}. This menace too, stands to worsen with the use of these video editing tools. What is required is a system that allows for users to gauge the trustworthiness of a video and displays this in an easy-to-understand manner. Of course, the scale of the problem and the heterogeneity of capturing devices and  underlying technologies means that no one-size-fits-all solution is going to be feasible; yet, that is no reason to try and find at least a one-size-fits-some solution to begin with.

In this paper, I present \emph{FrameProv}, a system for establishing end to end video provenance. Ensuring provenance represents a paradigm shift in the fight against fake news since it aims to allow ``ownership'' of videos and edits rather than the arms race of detecting malicious edits. At the core of FrameProv, is a novel yet intuitive data structure, the \emph{FrameChain}, which ensures integrity of videos at the individual frame level. Other major components required to make the entire system work are a video edits specification language, a (very) small trusted execution environment and a PKI model. I believe that FrameProv represents a first step towards a much-needed system of enabling trust in video streams.

While there have been several proposals for detecting forgeries~\cite{Cheng4284574, Kobayashi_detectingforgery, Wang:2007:EDF:1288869.1288876} as well as for embedding a trusted signal in the video feed~\cite{Hefeeda2015, Ou2016}, we haven't seen widespread adoption of any of these techniques. One of the reasons for this is the need for (legitimate) edits in video processing pipelines. Forgery detection and integrity mechanisms fail here because of the preponderance of false-positives -- they cannot discern the \emph{intent} behind the edits. FrameProv takes a different route: we start off with a simple, non-robust integrity measure on the raw video feed. Then, instead of trying to detect tampering of this feed, we allow the publisher to \emph{transparently} edit the video before publishing it. The edit information is then conveyed to the viewer who can then make a judgement call about whether the edits were justified or not. If the publisher tries to surreptitiously do any edits and not declare them, their video would fail the integrity test imposed by FrameProv. This is a paradigm shift in how we try to authenticate and assign trustworthiness to videos.

As we will see in Section~\ref{sec:litrev}, FrameProv is inherently different from most systems that preceded it not necessarily in the technology but in the aim of the technology. The name of the game used to be integrity, here it is \emph{provenance}, of everything including edits. Many of the design decisions I've made in designing this, however, have subtle knock-on societal effects at scale which make an open discussion about the system extremely pertinent (and interesting).

\section{Why Provenance?}
The core of the problem with video integrity is that edits are inevitable -- no one wants to watch an empty podium between speeches, there are very good reasons for wanting to have captions, annotations, etc. The reality of video publishing is that videos will go through multiple rounds of edits and compression. This is not a new problem; in countless other domains we recognise the need for edits and the accompanying necessity for establishing provenance for those edits.

One domain that most computer scientists will be most familiar with is code edits. Version control systems not only help with coordination on large code bases but also allow retrospective reviews of edits and the rationale and actors behind them. Another domain where provenance plays a crucial role is the legal system. In common law countries, precedents established in past cases help settle present disputes. More generally, old laws can be changed by the use of amendments which are, in essence, edits to the original text. In the EU, these amendments are tracked by noting down the date, rationale and presiding bodies behind the amendment (see~\cite{eu-amendment} for an example). Some countries go even further: in Australia, the Federal Register of Legislation records not only the amendments but also all the evidence and testimonies that went into making the amendment in order to record the entire lifecycle of laws~\cite{auzlegis}.

These systems are an affirmation of the necessity of edits to ``source'' data and the consequent need to be able to trust those edits down the line. FrameProv aims to introduce the same paradigm to videos.

\section{Related Work}
\label{sec:litrev}
As mentioned earlier, most of the related work in this field has been directed at detecting video forgeries. In this section, we will briefly go over the categories of these techniques and elaborate on how they differ from FrameProv.
\subsection{Video Forensics}
By far the most amount of related literature exists in the field of video forensics. These forensics fall under three broad categories: compression artefacts detection, scene processing and source detection. 

Compression artifacts have been widely studied~\cite{milani_overview_2012}. These seek exploit algorithm-specific artifacts to estimate the number and kinds of edits performed to a particular video. Most compression algorithms work on a block-by-block (where each block is an n x n set of pixels) basis, this is why many researchers have looked into these blocks as possible evidence for detecting recompression. Earlier models used to rely on comparing the statistical differences in pixels within the same block versus across block boundaries~\cite{PMID:18237903, MLE901117}. However, due to the introduction of deblocking filters (which try to smooth out compression artifacts) in modern video coding architectures, these methods have been rendered less effective and research is underway to improve upon their accuracy. Another vector that has been explored is detecting double quantization -- differences in noise levels in the image due to the use of multiple compression algorithms on select sections of the image. This method was first used to detect recompression artifacts in MPEG streams~\cite{Wang:2009:EDF:1597817.1597826}; this was then further developed to detect the codecs used in the first compression stage~\cite{VCI6288363}. However, despite the extensive research, double encryption detection remains a difficult problem because of the vast diversity of video coding architectures and all the possible combinations possible therein.

I have used ``scene processing'' as a catch-all phrase for techniques that infer semantics from the content of the video clips in order to detect tampering. These include things like detecting geometric irregularities in the scenes, hue estimation, shadow detection, etc. Here, I will highlight a sampling of interesting ideas. Perhaps the most generalisable idea is that of geometric irregularities -- this category includes ideas like detecting irregularities in illumination~\cite{Johnson_exposingdigital}, misaligned perspectives on object placed in different locations in the video~\cite{Signs5652906} and irregular trajectories of flying objects~\cite{ballistics5995165}. Another interesting idea is that of blur estimation: since many forgeries rely on blurring tampered regions or frames, the researchers devised a way for detecting artificially introduced blur regions~\cite{blur5504476}.

The last field of video forensics I would like to discuss is that of source or camera identification. The field was kick-started in~\cite{fingerprint817172} where the researchers sought to identify individual camcorders using unique patterns on a camera's CCD chip. Since then there have been several follow-ups to this original idea, with recent works expanding to fingerprinting using aberrations in camera lenses~\cite{lens2006SPIE.6069..172C} as well as a camera's colour filter array~\cite{cfa1530330}.

As stated earlier, the goal of forensics and FrameProv are quite different --  the goal of FrameProv is not to detect \emph{whether} there has been an edit or not but rather \emph{what} edit has been performed and whether it has been declared to the viewer. 

\subsection{Integrity Measures}
There has been extensive research into watermarking techniques~\cite{water7938666} and some of the ideas developed in that field have applications in forgery prevention/detection, however, overall I consider it to be a tangential strain of research.

In my review of the literature, I have found the most relevant research to FrameProv in patent filings rather than in academic papers. One of the earliest patents for integrity measures was~\cite{patent:7477749}. Here, the authors introduced a rudimentary system for DRM enforcement. This involved a central signing authority validating all the content by signing hashes of videos. \cite{patent:20080037783} refined this idea specifically for video streaming; here, the authors introduced authentication frames interjected between snippets of streamed content that validated the preceding snippet. This is very similar to what FrameChain does albeit not at the same level of granularity.

The piece of work closest to FrameProv in this domain is~\cite{patent:china}. Here, the system injects ``integrity measures'' within the video feed itself (just like in FrameChain); if the adversary tampers the video feed then the integrity measure will fail. Unfortunately, this is a Chinese patent and I have been unable to locate a good translation for this thus limiting my understanding of the system. Due to this limitation, it is also difficult for me to pinpoint the technical differences between their integrity measure and FrameChain. In any case, like all the measures mentioned in this subsection, this system too does not try to nuance the difference between legitimate and illegitimate edits.

\subsection{Related Image-based Techniques}
A tangentially related piece of work is ~\cite{Kee19907}. In this paper, the authors tackled the problem of retouching in fashion magazines. To quantify the extent of the retouching they came up with a metric that aligned closely with subjective opinions. While this project isn't tackling the same problem, the similarity it shares with FrameProv is that they are both trying to distinguish between legitimate and illegitimate edits rather than just a black-or-white edited/non-edited statement.

\section{A Provenance-revealing user interface}
\begin{figure*}
  \includegraphics[width=\textwidth]{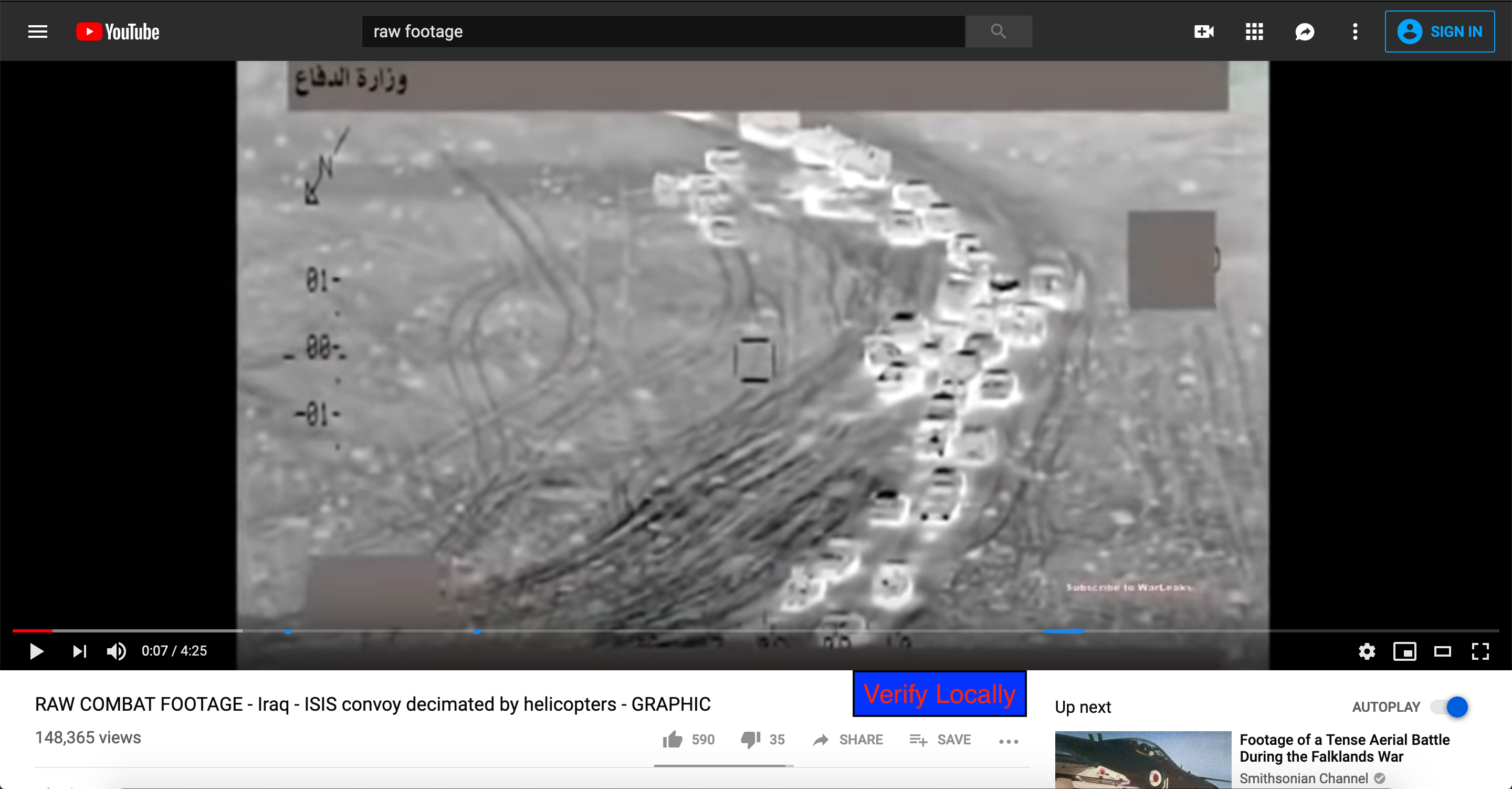}
  \caption{YouTube has trained users to associate yellow boxes on the timeline with ads. We could use this same mechanism to expose edits to the user. Here, we see a mock up of how blue boxes could be used to show edit locations. Hovering over the blue boxes shows edit information. The ``Verify Locally'' button allows the user to download raw provenance data and validate it locally without trusting the hosting platform.}
  \label{fig:youtube}
\end{figure*}
The make-or-break feature of this system is the user interface. If the users don't \emph{feel} that they can trust it, or if they can't understand it intuitively, then no amount of cryptography will help. Luckily, we have some assistance in this department. There are several evils that ubiquitous communication has brought with it, but there is one good that has accompanied it which we can utilize here: users now have very good mental models of what video streaming looks like and what video annotations mean. This is a major change in the landscape: instead of having to deal with myriad video players (and user expectations), we now only need to target one. Moreover, YouTube has also introduced timestameped annotations to the masses: everyone knows that the yellow lines on the YouTube seek bar denote advertisements. What if we could leverage this mental model to annotate provenance data of videos in an easy-to-understand manner?

Figure~\ref{fig:youtube} shows a mock-up of one such system. I am leveraging the user's expectation of seeing relevant information in the seek-bar to also serve edits information. Ads show up as yellow bars as before, edits show up as blue bars. Hovering over the blue bar shows the edits performed at that point. Additionally, we present a button to the user for locally verifying the provenance of the video; I feel that this option makes it easier to trust the interface than an opaque ``take-it-or-leave-it'' checkmark. This interface is the final user-facing product that we need to deliver regardless of the underlying infrastructure. The remainder of the paper is dedicated to discussing the various options available for said infrastructure.

\section{Infrastructure and Adversary Model}
\begin{figure*}
  \includegraphics[width=0.75\textwidth]{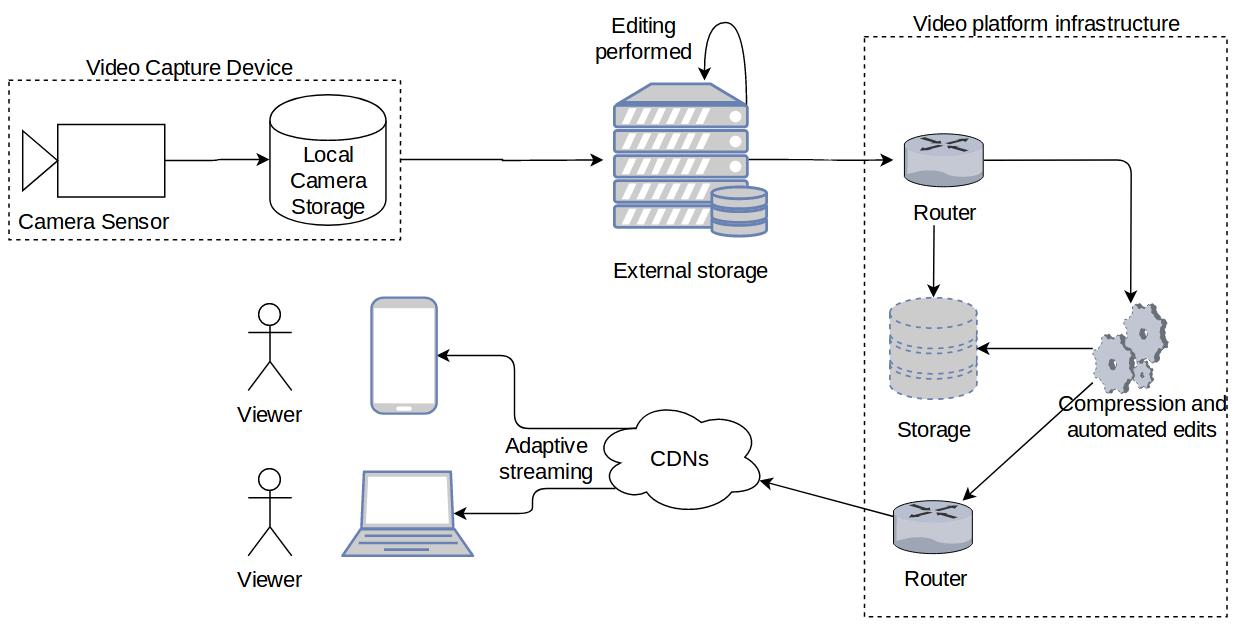}
  \caption{High level overview of a typical video transmission chain}
  \label{fig:vidtrans}
\end{figure*}
Figure~\ref{fig:vidtrans} shows an overview of the various steps involved in the broadcasting of videos. There are many more intermediary components but for the purposes of our discussion, this level of granularity will suffice. 

The first decision we need to make is to choose the ``ends'' of our end-to-end system. Ideally, we would want our system to cover the entire video transmission chain from the camera sensor to the viewer's screen. This minimises the security perimeter of the system to just the camera sensor and the display. With FrameProv, we come quite close to this ideal. I decided to embed a small trusted execution environment in between the camera sensor and the local camera storage. While this does introduce the need to trust some form tamper-evident packaging, I believe that this is the closest we can get to a true end-to-end system without designing custom camera sensors. 

On the other end of the transmission chain, I assume that the plugin used to verify the video (whether running on the viewer's computer or on a hosted platform) executes the verification script correctly. I also allow for variants that introduce a trusted editing platforms for ease of distribution of finished videos at the cost of trusting the platform. In addition to these two points of trust, we also need to introduce a trusted publicly viewable key store; I'll elaborate on this in the next section.

Given all of this, I've assumed a pessimistically powerful adversary model. We assume that the adversary has compromised the entire video transmission chain except for the trust perimeter around the sensor and the verification device. We assume that the viewing device is not compromised and performs verification of the feeds correctly (and gives appropriate notifications to the user). We assume that the adversary cannot inject anything between the camera sensor and the TEE, or tamper with the tamper-evident packaging in any way without leaving obvious evidence of the tampering. Lastly, standard cryptographic assumptions apply: the adversary cannot find collisions for hashes, cannot forge signatures, etc.

\section{System overview}
In this section I'll give a high level overview of FrameProv. First, I'll list the system requirements; then, I'll introduce the data structures I've used to carry the provenance information and lastly we'll look at the infrastructure and the various design choices therein.
\subsection{System requirements}
FrameProv must fulfil the following requirements
\begin{itemize}
    \item The ability to associate video feeds with specific camera sensors
    \item The ability to validate the integrity of a raw video feed
    \item Support for signed edits that can later be exposed to the users
    \item An easy-to-understand, intuitive user interface
\end{itemize}

\subsection{Data structures}
There are two main data structures necessary for the functioning of FrameProv: FrameChain and the Video Edit Specification Language (VESL). Let's now look at each in turn.
\subsubsection{Framechain Data Structure}
\begin{figure}
\label{fig:fc}
\centering
\includegraphics[width=\linewidth]{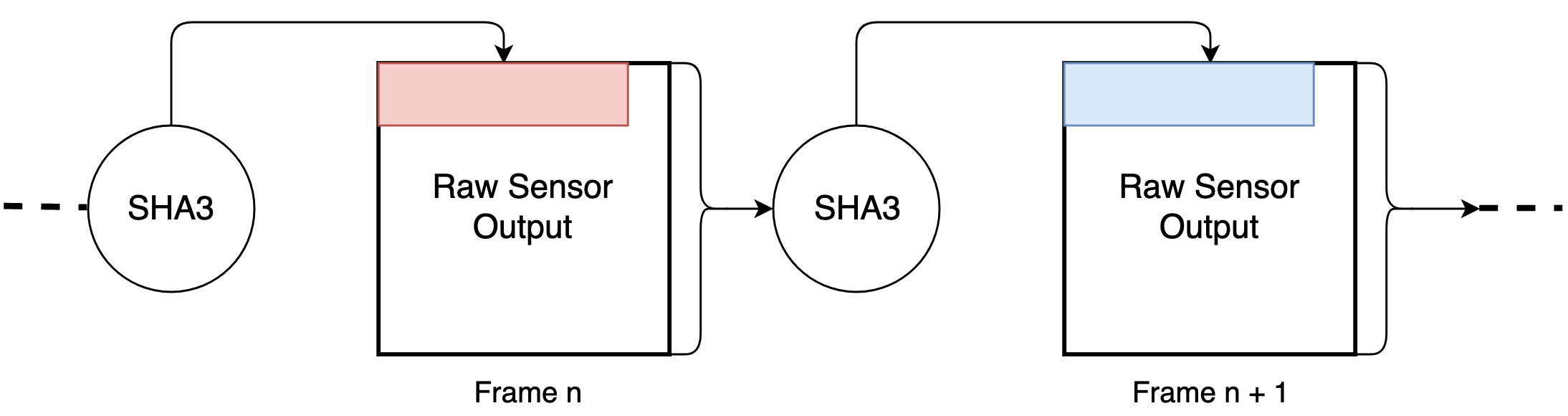}
\caption{The Framechain data structure}
\label{fig:fc}
\end{figure}

The framechain data structure is effectively a blockchain where the blocks are RGB 2D-arrays and the hashes are represented as a subset of the first row of RGB pixels. Figure~\ref{fig:fc} illustrates this idea. More specifically, a framechain file is a set of $n$ 2D arrays where the first array encodes metadata about the camera sensor and the Tiny TEE (which we'll see in the next section). Arrays $2$ to $n-1$ are raw RGB video frames where the first row of pixels encodes the hash of the last array. Array $n$ encodes a signature over the hash of array $n-1$ and, optionally, a set of other preceding arrays.

One might question the need for the framechain data structure. Why not simply hash the entire video file after it is recorded and sign that? The reasons for not going through with this simpler approach (and other similar strawmen approaches), briefly, are as follows:

\begin{itemize}

    \item \textbf{Streaming.} One of the advantages of the framechain approach is that it is possible to stream an authenticated video stream without knowing what the final hash would be.
    \item \textbf{Single data stream.} The framechain stream is self-contained. No external stream of data is required to authenticate the video feed. This simplifies the constraints on the supporting infrastructure as it is now possible to use any existing video streaming platform.
    \item \textbf{Visual indicators and ease of verification.} The rapidly varying row of pixels at the top of the video serve as a visual aid to the viewer, indicating that the video they are watching may use FrameChain provenance technology irrespective of the video player they may use to view it. This also makes designing verification plugins for existing video players an easier task since we are not dependent on container-specific metadata for the authentication stream; it is simply a matter of reading the values of a specific set of pixels.
    \item \textbf{Snippet signing.}  With FrameChain, it is possible to affirm the integrity of sections of the video feed; this could help with the editing process by reducing the number of declared edits during relatively unedited sections of the video. 
\end{itemize}

Verification of the framechain data structure (with no edits) is a simple matter of iterating over the feed from the first array to the last, verifying the hashes at each step along the way before finally verifying the signature encoded in the last frame. If multiple frames were signed in the last frame, then the verifier informs the user of the additionally verified snippets so that they can be used instead of the entire video feed.

\subsubsection{Video Edit Specification Language (VESL)}
The VESL is a JSON-like markup language that encodes edits to a framechain file. The VESL encoded edits are what allow the publisher to make their edits known to the viewer. That said, the particulars of the language aren't too crucial for the architecture of the system -- as long as the verification script can parse it, any language can be used to specify the edits. Every VESL file must be accompanied by a signature over it; this allows an editor to assert ownership over the edits.

As it stands, VESL only supports reductions, not additions. That is, we have ways to specify compression, frame drops, changes in playback speed, etc. However, we do not yet support operations such as overlaying text or adding external graphics: operations that require interfacing with additional sources of graphics. I acknowledge that these are crucial operations with legitimate uses; figuring out how to support these operations within VESL remains one of the major challenges with FrameProv. Listing 1 shows a simple VESL example that illustrates frame deletion, applying preset colour filters before applying compression to the entire file. To reduce verbosity of the VESL file, defaults from the widely used video editing software, FFmpeg~\cite{ffmpeg}, are used wherever they are unspecified within VESL.

\begin{lstlisting}[caption=VESL example]
{
    "editType" : "rangeDeletion",
    "rangeDeletionParams" : {
        "fromFrame" : "1250",
        "toFrame" : "1500"
    },
    "editType" : "videoFilter",
    "videoFilterParams" : [
        {
            "filterType" : "alpha",
            "fromFrame" : "2010",
            "toFrame" : "2020",
            "typeParams" : {}
        },
        {
            "filterType" : "atadenoise",
            "fromFrame" : "2040",
            "toFrame" : "2090",
            "typeParams" : {
                "0b" : "1.6"
            }
        }
    ],
    "editType" : "compression",
    "compressionParams" : {
        "algorithm" : "H.264"
        "algorithmParams" : {
            "CRF" : "27",
            "preset" : "veryfast",
            "twopass" : "true"
        }
    }
}
\end{lstlisting}

\subsection{Delivery Modes}
\begin{figure*}
  \includegraphics[width=\textwidth]{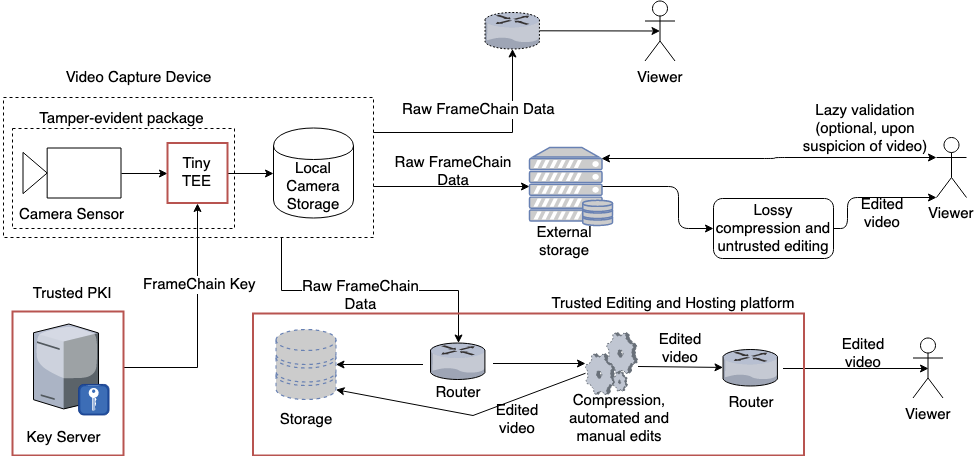}
  \caption{Overview of FrameProv highlighting the three different possible delivery modes. Components in red are trusted.}
  \label{fig:e2e}
\end{figure*}
One of the shortcomings of the FrameProv system as described above is that it operates on raw RGB arrays and it can only affirm the veracity of these arrays. Any kind of compression or editing would render the hashes invalid and would thus fail. This is an issue since for many applications where compression, at the very least, is necessary. To enable these operations, as illustrated in Figure~\ref{fig:e2e}, I envision three different delivery modes for framechain feeds.

\subsubsection{Direct stream}
The simplest delivery mode for FrameProv is direct streaming. The viewer downloads the raw framechain data and renders it locally on her machine. This allows for immediate validation of the video feed without introducing any additional trusted components. The drawback of this approach is the immense bandwidth requirement of downloading uncompressed frame data and the lack of any ability to do edits.

\subsubsection{Trusted editing platform}
Here we introduce a trusted actor, namely, a trusted editing platform to both give the means to edit the video as well as to host it for viewing. When an uploader uploads the raw framechain, the platform first verifies it and only allows the uploader to proceed if the verification succeeds. Next, it allows the uploader to compress the video and to do edits. The set of allowed edits is currently limited but I am looking into ways to expand this to support arbitrary edits. Then, the uploader can open the video to public viewing. Figure~\ref{fig:youtube} shows what the end result of this process would look like.

While this approach introduces a trusted intermediary, I believe that this is the most promising approach given the low bandwidth requirements on the servers as well as the end users.

\subsubsection{Delayed verification}
Conceptually this is quite similar to the trusted editing platform except that instead of trusting the editing platform, the verification is done by the user on their local machine. The flow in this case looks like this:
\begin{enumerate}
    \item Viewer downloads compressed and possibly edited video.
    \item After viewing the video, viewer decides to verify its integrity. She requests the source framechain data and the VESL-encoded edits. Each edit \emph{must} be accompanied by a valid signature (this identifies the editor).
    \item Viewer creates a local copy of the video by performing the VESL-encoded edits on the framechain data.
    \item Viewer hashes the computed video and the downloaded video and compares the hashes. If the hashes match, verification has succeeded.
\end{enumerate}

\subsubsection{Hybrid approach}
\label{subsub:hybrid}
Delayed verification can also be used as a means to mitigate the trust issues of the previous approach: the trusted editing and hosting platform can  provide users with the option to download the framechain and VESL files to perform local validation. This hybrid verification approach allows users on constrained devices (in terms of bandwidth or computational power) to view edited videos while allowing unconstrained viewers to verify the devices locally and keep the platform honest.

\subsection{Tiny TEE} Tiny TEE is a (very) small trusted component that only has one function: create a framechain from the raw sensor outputs. To do this, it needs to be able to perform three operations: hashing, signatures and initial frame generation. The sequence of operations are as follows:
\begin{enumerate}
    \item User presses ``record'' on the camera package.
    \item Tiny TEE generates a new initial frame encoding camera information, TEE hardware-based public key and other metadata (timestamp, sequence number, etc.)
    \item The sensor is activated and starts recording. The first frame from the sensor is sent to the Tiny TEE.
    \item Tiny TEE hashes the initial frame, encodes it as a row of pixels and embeds it onto the first sensor frame.
    \item This operation is repeated until the user presses the stop recording button.
    \item Upon receiving the stop signal, the sensor stops recording. Tiny TEE signs the hash of the last frame appended by video metadata (number of frames, framerate, etc.). This signed data and the signature are encoded as an RGB frame and sent to storage.
\end{enumerate}
Tiny TEE is co-located with the sensor in a tamper-evident package. Thus, if an adversary tries to inject fake feeds into the TEE, they would first have to destroy the packaging. Any later investigation would see this tampering and cast doubt over the footage.

One might wonder why I chose to go with tamper-evidence rather than tamper-resistance. This is because tamper resistance cannot be guaranteed. Getting into the device is simply a function of investment. To be certain of the veracity of the video feed, we would have to inspect the package anyway. Tamper-resistance does not afford us any additional security guarantees beyond (possibly) increasing the investment required by the adversary for an attack.

\subsection{FrameProv PKI}
\label{sub:pki}
Implicit in the discussion about Tiny TEE has been a PKI: we need something to supply the TEE with a hardware-based private key that is used to sign the FrameChain. This PKI requirement has been the weakness of every TEE mechanism proposed so far, and for good reason -- the entity managing the PKI potentially has the power to make arbitrary assertions about the existence of the TEEs and the operations performed by them. PKIs for TEEs are tricky, there's no getting around that fact.

However, I do believe that there are properties about our scenario that make it possible to come up with an acceptable solution. For one, device privacy is not a requirement. In fact, exposing the identity of the responsible TEE is a requisite feature not a bug. Existing TEE systems such as Intel SGX have had to use privacy preserving ID schemes~\cite{sgxepid} because of their need to hide the TEE identity. This leads to a situation where Intel could do sybil attacks by creating phantom TEEs. When we give up on the privacy requirement, we take away the problem to a large extent. Now, we can simply publish all the public keys generated by our PKI along with their genesis timestamps and even which organisation owns a particular TEE. This severely reduces the scope of possible sybil attacks that a malicious PKI can commit.

Another factor that helps us is the fact that our TEE is not a general purpose TEE. In fact, since it only ever interfaces with one component and only ever does one set of operations, it is possible to have a completely non-reprogrammable TEE. This reduction in scope allows us to do away with difficult primitives such as remote attestation. Simply verifying the signature is sufficient for validating the output of Tiny TEE.

Of course, despite these reductions in scope, the PKI still is a point of failure. We still need to trust it to delete the private key once it has been injected into the TEE. If it doesn't do so then it can create arbitrary FrameChain masquerading as having originated from that TEE. This is a trust issue that is hard to mitigate, despite some promising recent industrial solutions~\cite{securethingz}. Another interesting PKI question is how to assign identities to the edits that are declared within VESL; do we use a simple git-like sign-off declaration or try to establish a PKI? The answer to this is inherently tied to how we expect the videos to be published. 

\section{Deployment scenarios}
At present, I have implemented a prototype of FrameProv using a Sony IMX219 sensor and a raspberry pi simulating Tiny TEE. I have also implemented verification scripts that can be run on local FrameChain data (Python) as well as a Firefox plugin that can operate on remotely hosted data. This gives some confidence that such a solution is at least technically feasible. The difficult question then is how do we bring this system to the real world? How do we deploy FrameProv in a sensible manner that allows for scale while still maintaining public perception of trust in the system?

I do not envision FrameProv being available on commodity hardware to begin with. I think that the best way to deploy such a system in the field would be by first issuing it to trusted news networks and integrating the verification plugin into major video hosting websites (with the option to do local verification, as mentioned in section~\ref{subsub:hybrid}). I have had informal talks with two video platforms and there does seem to be a real interest in such a system. From their perspective, having an independent third party develop such a system reduces the amount of trust that needs to be placed in the video platforms. 

As a stop-gap solution, it is also possible to do the verification via a browser plugin. As mentioned above, I have created a prototype verification plugin which reads the descriptions of hosted videos for links to raw framechain and VESL data, loads it and then verifies the veracity of the displayed video. While this approach introduces the additional friction for users of installing a plugin, and for uploaders to host the raw feed data, it allows us to deploy FrameProv today without waiting for integration with major hosting platforms. This is an easy way for us to expand this proof-of-concept to the open Internet and refine the system before major integrations with video platforms.

\section{Should we do this?}
An interesting topic of discussion would be to ask ourselves: let's say we can actually deploy a large scale institutionally supported version of FrameProv, then should we? So far, we have listed the pros of having a FrameProv system to society, but there are genuine concerns as well.  In this section, we discuss these concerns and point at possible mitigation for each, if any.

\subsection{Citizen Journalism}
One way that FrameProv may end up doing more harm than good if it were to be publicly adopted is in the public perception of non-FrameProv videos, namely, that it may debase public trust in them. If we assume that FrameProv's Tiny TEE isn't widely incorporated into mobile phones\footnote{While there isn't any fundamental technical difference between camera sensors in phones and standalone cameras as far as difficulty of integration is concerned, there are organizational difficulties with the former. Phone companies tend to maintain a tight systems integration and convincing them to incorporate a new TEE onto the SoC would be challenging. A possible solution is to use existing TEEs available in phones~\cite{applesecureenclave, trusty} instead of Tiny TEE.} but primarily in dedicated cameras held by news organizations, this would lead to a situation where there is effectively a new trust tier that is only accessible to news organizations. 

Citizen journalism has seen a promising rise in recent years, driven by the ubiquity of phone cameras; what happens when those cameras are seen as fallible witnesses rather than evidence? Is the death of citizen journalism the price we have to pay in order to make society view videos with higher scepticism? There is an argument to be made that perceiving video footage as equivalent to witness testimony is overall a positive thing as it gets at the heart of the fake news problem; however, this shift would also do away with all the good brought about by a ``sousveillance'' world~\cite{transparentsociety}. This is therefore a decision that needs to be made with open eyes and after careful deliberation, not sleep-walked into. Indeed, the ideal situation from a citizen journalism point of view would be where FrameProv is ubiquitously deployed. This, however, may lead to other concerns as we shall discuss in the following subsections.

\subsection{PKI Concerns}
Another reason why FrameProv might not be such a good idea is because it \emph{potentially} places the keys to trust on the Internet in the hands of the maintainer of the FrameProv PKI. I have discussed the technical reasons for why this doesn't necessarily have to be a deal-breaker in Section~\ref{sub:pki} but it still leaves open the question of who operates the PKI and TEE manufacturing facility. The answer to this is a function of economics, politics and public perception more than technical expertise. FrameProv is a system for the benefit of a fixed set of news organizations, and theoretically, as long as these organizations can agree upon a custodian for the PKI (along with the necessary oversight) most of the issues of centralization can be abstracted away. However, the real world of alliances is messy and public opinion on such matters can be a fickle thing. This raises interesting questions about public-facing custodian schemes that we haven't quite had to face at scale yet.

Similar concerns apply, albeit to a less severe extent, to the signatures accompanying VESL-encoded edits. These signatures would ideally be mapped to the video editors in order for the viewer to have an insight into the entire provenance chain of a video. However, this makes it very important to be able to trust the PKI infrastructure since editors would be staking their reputations on it.

\subsection{Privacy Concerns}
\label{sub:privacy}
Privacy concerns arise in three ways: preserving the privacy of subjects in the video, preserving the privacy of the video recorder and preserving the privacy of the video editors. The first of these requires us to re-think the verification process. Let's reason about this with an example. Let's say we are using FrameProv to record a whistleblower's testimony which includes her appearing in front of the camera. In this situation, we want to be able to assert that we have made no tweaks to the recording but we also want to preserve the whistleblower's anonymity. This, necessarily, means that we cannot release the raw FrameChain data to public at large. The next best thing we can do, and how journalists actually tend to work, is to \emph{escrow} the trust. So, instead of releasing the raw files to all viewers, we get in trusted third parties to view the raw files and validate the FrameChain. Once they do that, they can then assert that the edited video (with, say, black bars over the whistleblower's face) is not misleading. Then, we can use this asserted video as the new root of trust and make it available for public viewing. Such a trusted third party arrangement may prove to be necessary for a number of sensitive videos.

Preserving the privacy of the video recorder entails a trade-off: being able to pinpoint the recording device allows us to verify the integrity of the TEE but exposes the identity of the recorder (or the owner of the device). This can be made less severe by using privacy preserving group signature schemes, as mentioned in Section~\ref{sub:pki}, and making TEE verification voluntary. Thus, the recorder of the video gets to choose whether or not to reveal their identity. Preserving the privacy of the video editors is even simpler: the editors can simply use ephemeral keys and throw them away after signing the VESL files without putting them on any look up infrastructure. In this way, the keys end up serving as ephemeral pseudonyms. Of course, this would mean that the viewer will see edits being made by an unidentified editor which may affect their trust perceptions of the video.

\section{Future Work}
While FrameProv is still in the very early stages of development, it is quite clear already that any such system requires extensive related research to determine the consequences of their deployment. Here, I list a few such pieces of future work.

\subsection{Sensible defaults for edit warnings}
Experience has shown that warnings are only effective when they are seen as useful. This means that we need to be selective in which edits we show most prominently to the end user; the decision of which those should be is not immediately obvious to us. Clearly, cutting out of frames should be remarked upon but the use of compression can probably be conveyed more subtly. We do not want to overwhelm the user with warnings nor do we want them to miss out on a critical edit -- this tradeoff represents a challenging, and interesting, HCI problem.

\subsection{Anchoring to blockchains}
There are several ways in which we can utilize public blockchains to further increase the integrity guarantees of FrameProv. The most intuitive would be to time constrain the video: if the recording device has access to the state of, say, the Bitcoin blockchain~\cite{Nakamoto_bitcoin:a}, it can retrieve the latest mined block. If the device does this before recording the video and encodes the hash of the block in the first frame then we can prove that the video must have been recorded after that block was mined. One may choose to use non-forking blockchains to get tighter timing constraints on this assurance. After the video has been recorded, the recorder can also hash the FrameChain and anchor that hash on to a public blockchain by sending it as a transaction payload. This helps us prove that the video must have been recorded before that block was mined. Thus, in combination, these two anchors give us a time frame for when the recording took place. This gives us another datapoint to increase our trust in the video. If all editors do this anchoring then we can have a trusted timestamp to go with the entire video provenance chain making tampering even more difficult.

Another way in which we can leverage public blockchains is by the look up details of the editors on the blockchain. While this introduces concerns around revocation and data regulations, it would help allay fears of foul-play by any centralized look-up service.

\subsection{Re-recording detection}
One attack that could be launched on FrameProv is the following: attacker records a video, maliciously edits it and then plays it back on a screen. Then the attacker records the edited video with a FrameProv device. Such a video would pass the verification process. There has been research into detecting such replays in the context of copyright protection~\cite{Jung2015}, however, this needs to be expanded and included into the verification script itself.

\subsection{Edit-heavy videos}
The discussion surrounding the usability and delivery of the FrameProv system has so far made an implicit assumption about the kinds of videos we are interested. The system proposed as is works best for short videos with minor edits and single source file. However, that covers a small (if important) subset of videos in general. Many videos use multiple source files spliced together -- this would require us to come up with a way to meaningfully show the provenance of multiple files in a final video. It is also not uncommon for editors to throw away the vast majority of the recorded content in the final video -- this would force us to reconsider how we display edits; perhaps in this case it makes more sense to show what has been retained rather than what has been pruned. This last use case also raises questions about storage and bandwidth requirements: raw video files are massive, and if we need to store many times the length of the final video then is it even feasible to do verification? Such concerns give another reason for us to further develop the escrowed trust models presented in Section~\ref{sub:privacy}.

%\balance
\section{Conclusion}
Videos are the next frontier for the fakesters to exploit due to the emergence of new fabrication techniques. Existing literature has focused on detecting forgeries or on identifying the source cameras in order to do forensics. These tools, however, are woefully inadequate due to the necessity of legitimate edits as well as the mass distribution of videos on modern video platforms that make custom forensics infeasible. In this paper, I have presented an alternative paradigm -- instead of trying to detect or prevent changes to the source video, we should allow the editors to declare all of their edits to the viewer. In order to enable this paradigm, we introduced FrameProv, a prototype system that creates signed source files that can be edited in a verifiable way by the end viewer. I believe that this idea of allowing declared edits while disallowing surreptitious ones is the way forward in dealing with deepfake and fraudulent videos.

\section*{Acknowledgements}
I would like to thank the attendees of NSPW 2019 and both my pre-proceedings and post-proceedings shepherds who gave me invaluable feedback on this project. Their insights have made me think much more broadly about the context within which FrameProv finds itself and, I hope, this paper reflects that broader viewpoint. My PhD is supported by a Thales e-Security academic scholarship.
%
% The next two lines define the bibliography style to be used, and the bibliography file.
\bibliographystyle{ACM-Reference-Format}
\bibliography{sample-base}

\end{document}